\documentclass[twocolumn,showpacs,prl]{revtex4}
\usepackage{graphicx}
\usepackage{dcolumn}
\usepackage{bm}

\begin{document}

\title{General Hubbard model for strongly interacting fermions in an optical
lattice \\
and its phase detection}
\author{L.-M. Duan}
\affiliation {FOCUS center and MCTP, Department of Physics,
University of Michigan, Ann Arbor, MI 48109}

\begin{abstract}
Based on consideration of the system symmetry and its Hilbert
space, we show that strongly interacting fermions in an optical
lattice or superlattice can be generically described by a lattice
resonance Hamiltonian. The latter can be mapped to a general
Hubbard model with particle assisted tunneling rates. We
investigate the model under population imbalance and show the
attractive and the repulsive models have the same complexity in
phase diagram under the particle-hole mapping. Using this mapping,
we propose an experimental method to detect possible exotic
superfluid/magnetic phases for this system. \pacs{{03.75.Ss,
05.30.Fk, 34.50.-s}}
\end{abstract}

\maketitle

Among the control techniques for ultracold atoms, optical lattice and
Feshbach resonance play particularly important roles. The optical lattice is
used to control the interaction configuration while the Feshbach resonance
is a tool to tune the interaction magnitude. The combination of these two
powerful techniques naturally becomes the next frontier, which has attracted
significant recent interest \cite{1,2,3,4,5}. To understand this important
system, one needs to have a Hamiltonian to describe strongly interacting
atoms in an optical lattice. The starting Hamiltonian is unfortunately
complicated as one has to take into account multi-band populations as well
as direct neighboring couplings \cite{2,4,5}. We have described a method in
\cite{5} to derive an effective lattice Hamiltonian for this system from
field theory of the two-channel model.

In this paper, we report the following advance along this direction:
firstly, based on consideration of the system symmetry and its Hilbert
space, we show that a lattice resonance model turns out to be a generic
Hamiltonian for this system. The resulting Hamiltonian agrees with the one
from our previous microscopic derivation \cite{5}, but the method used here
shows this Hamiltonian should have general applicability. As an example, we
point out that for strongly interacting fermions in optical superlattices,
the effective Hamiltonian is again described by this lattice resonance model
when we introduce some dressed degrees of freedom. For certain
configurations of the superlattice, the system naturally supports d-wave
superfluid. Secondly, we mathematically map the lattice resonance
Hamiltonian to a general Hubbard model (GHM) with particle assisted
tunneling rates. The particle assisted tunneling brings in some new feature,
in particular, it may favor a superfluid phase compared with the Hubbard
model. Thirdly, we discuss the attractive Hubbard model with population
imbalance between the two spin components, and show it has the same
complexity in phase diagram as the repulsive Hubbard model under a
particle-hole mapping. This result is related to the recent large effort to
understand the polarized fermi gas \cite{6,7}. Finally, using the mapping
above, we propose an experimental scheme to detect possible exotic
superfluid or magnetic orders in this system. The method is based on
Raman-pulse-assisted time-of-flight imaging, and can reveal the superfluid
or magnetic phases with detailed information about the order parameter or
the pairing wave function.

For strongly interacting two-component (effectively spin-$1/2$) fermions in
an optical lattice, when two atoms with different spins come to the same
site, they form a dressed molecule with atomic population distributed over
many lattice bands due to the strong on-site interaction \cite{4,5}. We
consider the system with an average atom filling number $\overline{n}\leq 2$%
. In this case, we can neglect the $3$-atom occupation of a single site as
that is suppressed at low temperature by an energy cost about the lattice
band gap \cite{8}. We then have only four possible configurations for each
site $i$, either empty, or a spin-$\sigma $ ($\sigma =\uparrow ,\downarrow $%
) atom, or a dressed molecule. The creation operators for these
configurations are denoted by $b_{i}^{\dagger },a_{i\sigma }^{\dagger
},d_{i}^{\dagger }$, respectively, while the corresponding states are
written as $\left| b\right\rangle _{i}$, $\left| \uparrow ,\downarrow
\right\rangle _{i},\left| d\right\rangle _{i}$. We introduce the slave boson
operator $b_{i}^{\dagger }$ for an empty site $i$ so that the constraint of
the Hilbert space on each site can be simply implemented through
\begin{equation}
b_{i}^{\dagger }b_{i}+a_{i\uparrow }^{\dagger }a_{i\uparrow }+a_{i\downarrow
}^{\dagger }a_{i\downarrow }+d_{i}^{\dagger }d_{i}=I.
\end{equation}
Note that with this constraint, $a_{i\sigma }$ describe fermions while $d_{i}
$ and $b_{i}$ both represent hard-core bosons.

We assume the system has a global SU($2$) symmetry for the spin
components. In that case, $\left| \uparrow \right\rangle _{i}$ and
$\left| \downarrow \right\rangle _{i}$ are degenerate in energy,
and the most general form of the single-site Hamiltonian can be
written as $H_{i}=-\mu \sum_{\sigma }a_{i\sigma }^{\dagger
}a_{i\sigma }+(\Delta -2\mu )d_{i}^{\dagger }d_{i}$, where we have
absorbed the single-atom energy into the definition of the
chemical potential $\mu $, and $\Delta $ is the relative energy
shift of the dressed molecule. For two neighboring sites $i$ and
$j$, due to the atomic tunneling and off-site interactions, there
will be a Hamiltonian term $H_{ij} $ to describe all the possible
configuration tunneling or couplings. With the spin SU(2) symmetry
and the number conservation of each spin component,
the most general two-site Hamiltonian can be written as $%
H_{ij}=H_{ij}^{(1)}+H_{ij}^{(2)}$, where $H_{ij}^{(1)}$ describes the
configuration tunneling that involves transfer of one atom with the
following form (see the illustration in Fig. 1)
\begin{eqnarray}
H_{ij}^{(1)} &=&\sum_{\sigma }\left( ta_{i\sigma }^{\dagger
}b_{i}b_{j}^{\dagger }a_{j\sigma }+t_{da}d_{i}^{\dagger }a_{i\sigma
}a_{j\sigma }^{\dagger }d_{j}\right)   \nonumber \\
&+&g(d_{i}^{\dagger }b_{j}+d_{j}^{\dagger }b_{i})(a_{i\uparrow
}a_{j\downarrow }-a_{i\downarrow }a_{j\uparrow })+H.c.,
\end{eqnarray}
and $H_{ij}^{(2)}$ describes the configuration coupling that involves real
or virtual tunneling of two atoms with the general expression
\begin{eqnarray}
H_{ij}^{(2)} &=&\left( t_{d}d_{i}^{\dagger }b_{i}b_{j}^{\dagger
}d_{j}+H.c.\right) +x_{d}n_{di}n_{dj}  \nonumber \\
&+&x_{a}n_{i}n_{j}+x_{s}\mathbf{s}_{i}\cdot \mathbf{s}_{j}+x_{b}n_{bi}n_{bj}.
\end{eqnarray}
In $H_{ij}^{(2)}$, the number and the spin operators are defined by $%
n_{di}\equiv d_{i}^{\dagger }d_{i}$, $n_{i}\equiv a_{i\uparrow }^{\dagger
}a_{i\uparrow }+a_{i\downarrow }^{\dagger }a_{i\downarrow }$, $n_{bi}\equiv
b_{i}^{\dagger }b_{i}$, and $\mathbf{s}_{i}\equiv \sum_{\sigma \sigma
^{\prime }}a_{i\sigma }^{\dagger }\mathbf{\sigma }_{\sigma \sigma ^{\prime
}}a_{i\sigma ^{\prime }}/2$ ($\mathbf{\sigma }_{\sigma \sigma ^{\prime }}$
is the Pauli matrix). The term $n_{bi}n_{bj}$ is equivalent to the cross
coupling $n_{di}n_{j}+n_{i}n_{dj}$ under the constraint (1). By analyzing
the level configurations in Fig. 1, one can convince oneself that $%
H_{ij}^{(1)}$ and $H_{ij}^{(2)}$ include all the possible two-site coupling
terms with the SU(2) symmetry. As the atomic interactions are short-range,
all the multiple site couplings can be neglected. So a generic lattice
Hamiltonian is given by $H=\sum_{i}H_{i}+\sum_{\left\langle i,j\right\rangle
}\left( H_{ij}^{(1)}+H_{ij}^{(2)}\right) $, where $\left\langle
i,j\right\rangle $\ denotes neighboring sites. This Hamiltonian describes
the coupling between the fermionic atoms $a_{i\sigma }$ and the bosonic
dressed molecules $d_{i}$ with a detuning $\Delta $, and will be referred in
the following as the lattice resonance model.

\begin{figure}[tbph]
\centering
\includegraphics[height=3cm,width=8cm]{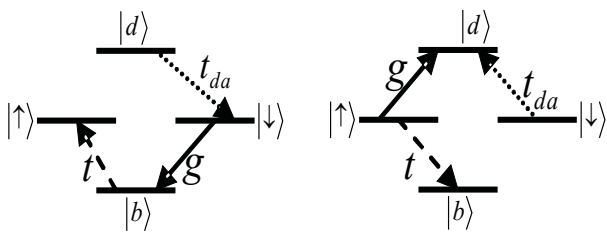}
\caption{ Illustration of the configuration tunnelling between two
neighing sites. The process shown in the figure correspond to the
$t$, $g$, and $t_{da}$ terms in the Hamiltonian.} \label{Fig1}
\end{figure}

The Hamiltonian $H$, together with the constraint (1), poses a well-defined
problem. Note that $H$ agrees in form with the effective lattice Hamiltonian
for strongly interacting fermions that we derived before from a completely
different method \cite{5}. The only specification from that microscopic
derivation is to fix the coefficients $x_{b}=0$\ and $x_{a}=-x_{s}/4$. As
mentioned in \cite{5}, in the case of a large detuning $\Delta $, the
Hamilton $H$\ is reduced to either the t-J model for atoms or the XXZ model
for dressed molecules, depending on which species get populated. We also
notice that for short range interactions, with increase of the lattice
potential barrier, all the interaction coefficients in $H_{ij}^{(2)}$ decay
much faster compared with those in $H_{ij}^{(1)}$. So for a lattice with
sufficient depth, $H_{ij}^{(1)}$ dominates over $H_{ij}^{(2)}$, and in the
following, without special mention we will consider the simplified
Hamiltonian $H=\sum_{i}H_{i}+\sum_{\left\langle i,j\right\rangle
}H_{ij}^{(1)}$ by dropping $H_{ij}^{(2)}$.

We now recast the Hamilton $H$ into a different form which shows its
connection with the Hubbard model. For this purpose, we map the dressed
molecule state $d_{i}^{\dagger }\left| vac\right\rangle $ to the two-fermion
state $a_{i\downarrow }^{\dagger }a_{i\uparrow }^{\dagger }\left|
vac\right\rangle $, where $\left| vac\right\rangle $ denotes the vacuum.
Note that physically the structure of the dressed molecules should be
determined by diagonalizing the on-site interaction Hamiltonian, and it
generally involves superposition of atoms in many band configurations \cite
{4,5}, which is certainly different from the state $a_{i\uparrow }^{\dagger
}a_{i\downarrow }^{\dagger }\left| vac\right\rangle $ with double occupation
on a single band. But mathematically we can identify these two states by a
one-to-one mapping. After this mapping, the Hamiltonian $H$ can be written
in the form
\begin{eqnarray}
H &=&\sum_{i}\left[ \left( \Delta /2\right) n_{i}\left( n_{i}-1\right) -\mu
n_{i}\right]  \\
&+&\sum_{\left\langle i,j\right\rangle ,\sigma }\left[ t+\delta g\left( n_{i%
\overline{\sigma }}+n_{j\overline{\sigma }}\right) +\delta tn_{i\overline{%
\sigma }}n_{j\overline{\sigma }}\right] a_{i\sigma }^{\dagger }a_{j\sigma
}+H.c.  \nonumber
\end{eqnarray}
where $\delta g\equiv g-t$, $\delta t\equiv t_{da}+t-2g$ and $n_{i\overline{%
\sigma }}\equiv a_{i\overline{\sigma }}^{\dagger }a_{i\overline{\sigma }}$ ($%
\overline{\sigma }=\downarrow ,\uparrow $ for $\sigma =\uparrow ,\downarrow $%
). To verify the two forms of $H$ in Eqs. (2) and (4) are equivalent to each
other, one can check the physical process represented by each term to
confirm it is identical. Note that in this new form of $H$, there is no need
of the slave boson operator to constraint the Hilbert space as the latter is
automatically fixed by the properties of fermions. As there is no additional
constraint, the Hamiltonian in the form of Eq. (4) looks simpler and may be
easier for treatment in certain cases. Compared with the conventional
Hubbard model, the effective tunneling rate in $H$ becomes an operator which
depends on occupation of the two sites. The original lattice resonance
Hamiltonian in Eq. (2) is thus mapped to a general Hubbard model ((GHM) with
particle assisted tunneling rates. For weakly interacting fermions, the
multi-band population and the direct neighboring coupling become negligible,
then the coefficients $g$ and $t_{da}$ tend to $t_{a}$, and the GHM\ returns
to the conventional Hubbard model as one expects in this case \cite{9}.

The derivation of the Hamiltonian $H$ in this work is based on very general
arguments about the single-site Hilbert space and the system symmetry. This
reminds us that $H$ has a generic form which should apply to different
systems with similar Hilbert space structure and symmetry properties. As an
example, we point out that for interacting fermions in an optical
superlattice, under several interesting configurations, the system is also
well described by the above Hamiltonian $H$. Figure 2A illustrates an
optical superlattice potential which can be realized with two standing wave
laser beams \cite{10}. With a combination of this superlattice and the
conventional optical lattice potentials, one can realize the dimer or
plaquette lattices as illustrated in Fig. 2B and 2C where the intra-dimer
(intra-plaquette) couplings are much stronger than the inter-dimer
(inter-plaquette) couplings. To derive an effective Hamiltonian for this
system, one needs to first construct dress energy levels for each dimer
(plaquette) by exactly solving a few-site problem. For two-component
interacting fermions in those lattices near half filling, the low energy
level configurations from each dimer (plaquette) have basically the same
structure as those shown in Fig. 1 \cite{11,12,note1}, and the system also
has the SU(2) symmetry. We then immediately conclude that the Hamiltonian in
the forms of Eq. (2) or (4) should be applicable to describe physics in the
dimer or plaquette lattices around half filling. The two-dimensional
plaquette lattice is particularly interesting: because of the internal
plaquette structure, the excitation from $\left| b\right\rangle _{i}$ to $%
\left| d\right\rangle _{i}$ states in Eq. (2) has a $d$-wave symmetry (e.g.,
$\left\langle d_{i}^{\dagger }b_{i}\right\rangle $ flips sign under a $\pi /2
$ rotation of the lattice) \cite{11,12}. When the effective detuning $\Delta
$ is tuned to be negative, one can show under pretty well controlled
approximations that the plaquette lattice in this configuration supports $d$%
-wave superfluid \cite{11,12}.

\begin{figure}[tbph]
\centering
\includegraphics[height=3cm,width=8cm]{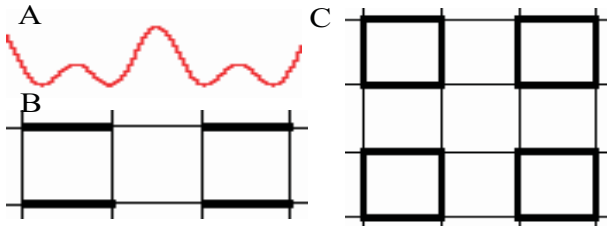}
\caption{ Illustration of an optical superlattice: (A) The superlattice
potential. (B,C) The dimer and the plaquette lattices (bold lines represent
stronger coupling) formed with the potential in (A).}
\label{Fig1}
\end{figure}

We now investigate some properties of the GHM\ in Eq. (4). When the detuning
$\Delta $ is negative, similar to the Hubbard model, we expect this
Hamiltonian is in a superfluid state away from the unit filling. When $%
\Delta $ is positive, although it is not clear yet whether $H$ has a
superfluid state, compared with the corresponding repulsive Hubbard model,
we do expect that the superfluid possibility becomes higher when $g>t_{a}$
(which is very likely the case for fermionic atoms near a wide Feshbach
resonance \cite{5}) as a large $g$ term (see Eq. (2)) clearly favors Cooper
pairing. From the single-site physics, we know that two atoms always have an
on-site bound state (corresponding to a negative $\Delta $) with the binding
energy ($-\Delta $) approaching zero as one moves to the BCS side of the
Feshbach resonance \cite{4,5}. So in the ground-state configuration, the
strongly interacting fermi gas naturally implement the GHM with a negative $%
\Delta $. To experimentally investigate the GHM\ with a positive
$\Delta $, one needs to start with the population in atoms
(instead of Feshbach molecules), and to approach the Feshbach
resonance from the BEC\ side (the system is in a metastable state
in this case). The effective Hamiltonians in different regions are
shown in Fig. 3.

\begin{figure}[tbph]
\centering
\includegraphics[height=3cm,width=8cm]{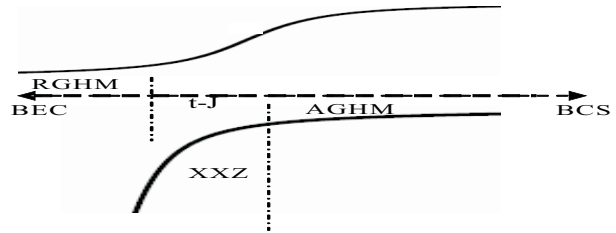}
\caption{ The effective Hamiltonians in different regions. The
solid curved correspond to two dressed molecule bands, and the
middle dashed line is an atomic band. On the BCS or BEC (with
population mainly in atoms) sides, the Hamitonians are given by
the attractive (repulsive) general Hubbard models (AGHM and RGHM),
respectively. As one increases the detuning $\Delta$, one gets
either the XXZ model for the dressed molecules or the t-J model
for the atoms. In the deep BCS or BEC limit, the GHM returns to
the attractive or repulsive Hubbard model, and the XXZ Hamiltonian
yields to the bosonic Hubbard model for molecules when multiple
occupation of a single site is allowed by weaker effective
interaction.} \label{Fig1}
\end{figure}

When we take into account possible population imbalance between the two spin
components, the repulsive and the attractive GHMs\ (with positive or
negative $\Delta $, respectively) become intrinsically connected, and they
should have the same complexity in phase diagram. Polarized fermi gas
recently raised a lot of interest \cite{6}, and in free space (or in a weak
trap), although population imbalance yields some new features, the basic
physics there is still largely captured by an extension of BCS\ type of
mean-field theory \cite{7}. However, for polarized fermi gas in an optical
lattice, we show that simple extensions of the BCS theory are very likely to
give misleading results because of the exact mapping between the repulsive
and the negative GHMs. Population imbalance corresponds to introduction of
an effective magnetic field $h$, which adds a term $-h\sum_{i}\sigma _{i}^{z}
$ ($\sigma _{i}^{z}\equiv n_{i\uparrow }-n_{i\downarrow }$) to the
Hamiltonian $H$ in Eq. (4). We apply a particle-hole transformation $%
a_{i\uparrow }\rightarrow a_{i\uparrow }$ and $a_{i\downarrow }\rightarrow
(-1)^{i}a_{i\downarrow }^{\dagger }$ to the Hamiltonian \cite{13} (for
simplicity, we consider a bi-partite lattice). Under this transformation,
the Hamiltonian $H-h\sum_{i}\sigma _{i}^{z}$ is mapped to

\begin{eqnarray}
&&H^{\prime }=\sum_{i}\left[ -\left( \Delta /2\right) n_{i}\left(
n_{i}-1\right) -\mu ^{\prime }\sigma _{i}^{z}-h^{\prime }n_{i}\right]  \\
&&+\sum_{\left\langle i,j\right\rangle ,\sigma }\left[ t_{\sigma }+\delta
g_{\sigma }\left( n_{i\overline{\sigma }}+n_{j\overline{\sigma }}\right)
+\delta tn_{i\overline{\sigma }}n_{j\overline{\sigma }}\right] a_{i\sigma
}^{\dagger }a_{j\sigma }+H.c.  \nonumber
\end{eqnarray}
where the parameters $\mu ^{\prime }\equiv \mu -\Delta /2$, $h^{\prime
}\equiv h-\Delta /2$, $t_{\uparrow }\equiv t$, $t_{\downarrow }\equiv t_{da}$%
, $\delta g_{\sigma }\equiv g-t_{\sigma }$, and we neglect the constant
energy per site $h-\mu $. One can see that an attractive GHM ($\Delta <0$)
is mapped exactly to a dual repulsive model (with $-\Delta $), where the
chemical potential $\mu $ and the filed $h$ exchange their roles. Superfluid
phases (including both the BCS state and the LOFF
(Larkin-Ovchinnikov-Fulde-Ferrel) state with pairing at nonzero momenta \cite
{14}) of the original model correspond to magnetic phases of the dual model
and vice versa. For a repulsive Hubbard model on a square lattice, the
magnetic order exists only in a region near half filling, and with hole
doping there are possibilities of exotic phases including a non-BCS\
superfluid state. This suggests for the attractive Hubbard model with
population imbalance, the superfluid phase exists in the region with small
polarization. With further increase of the polarization, there could appear
exotic phases including a d-wave magnetic order. Experimental investigation
of the attractive GHM\ with population imbalance (which might be easier for
realization compared with the repulsive one) is therefore able to provide
critical information to understand the challenging phase diagram of the
repulsive Hubbard model.

Finally, we propose a method to detect possible exotic phases in
this system by making use of the above mapping. Our purpose is to
directly measure the magnetic or superfluid order parameters. The
detection scheme combines the time-of-flight imaging with some
instantaneous Raman pulses \cite{15}. We take $^{6}$Li atoms as a
typical example. The scheme is illustrated in Fig. 4. Right after
turn-off of the trap, we immediately apply two consecutive
impulsive Raman pulses. These pulses are assumed faster than the
system dynamics (characterized by the Fermi energy), but slower
compared
with the level splitting between the $\left| \uparrow \right\rangle $ and $%
\left| \downarrow \right\rangle $ levels (about $70$ MHz). The first is a $%
\pi $-pulse, consisting of two laser beams propagating along different
directions, which transfers the atoms from the level $\left| \uparrow
\right\rangle $ to $\left| 6\right\rangle $ by imprinting a photon recoil
momentum $-\mathbf{q}.$ As the level $\left| 6\right\rangle $ is detuned
from $\left| \uparrow \right\rangle $ by a few GHz, this transition at the
same time tune the system out of Feshbach resonance (the atoms in states $%
\left| 6\right\rangle $ and $\left| \downarrow \right\rangle $ are only
weakly interacting) \cite{15}. The second is a $\pi /2$ pulse from
co-propagating laser beams applied to the levels $\left| 6\right\rangle $
and $\left| \downarrow \right\rangle $, which induces a transformation $a_{%
\mathbf{k}6}\rightarrow (a_{\mathbf{k}6}+e^{i\varphi }a_{\mathbf{k}%
\downarrow })/\sqrt{2}$ and $a_{\mathbf{k}\downarrow }\rightarrow (a_{%
\mathbf{k}\downarrow }-e^{-i\varphi }a_{\mathbf{k}6})/\sqrt{2}$ that
preserve the momentum $\mathbf{k}$ ($\varphi $ is the relative laser phase).
After these two pulses, we take the time-of-flight images (with basically
ballistic expansion) for the atoms in levels $\left| 6\right\rangle $ and $%
\left| \downarrow \right\rangle $, and the difference of these two images
give exactly the cross correlation of the $\uparrow $ and $\downarrow $
spin-components at different momenta:
\begin{equation}
n_{\mathbf{k}6}-n_{\mathbf{k}\downarrow }=2\mathop{\rm Re}\nolimits\left(
e^{i\varphi }a_{\mathbf{k+q,}\uparrow }^{\dagger }a_{\mathbf{k}\downarrow
}\right) .
\end{equation}

We now show through a few examples that we can directly confirm various
magnetic or superfluid phases with this detection ability. (i) For magnetic
phases with a pretty general form of the spin order parameter  $\left\langle
\mathbf{s}_{i}\right\rangle =\mathbf{v}_{1}\cos (\mathbf{Q}\cdot \mathbf{r}%
_{i})+\mathbf{v}_{2}\sin (\mathbf{Q}\cdot \mathbf{r}_{i})$ \cite{13}, we can
confirm it with sharp peaks for the correlation in Eq. (6) when the relative
momentum $\mathbf{q}$ is scanned to $\pm \mathbf{Q}$ . The spin vectors $%
\mathbf{v}_{1}$ and $\mathbf{v}_{2}$ can be inferred from the relative laser
phase $\varphi $. (ii) For the LOFF\ superfluid state with pairing at a
non-zero momentum $\mathbf{q}$, the order parameter $\left\langle a_{\mathbf{%
k+q,}\uparrow }a_{-\mathbf{k}\downarrow }\right\rangle $ is nonzero. After
the particle-hole mapping, this order parameter corresponds to a magnetic
order $\left\langle a_{-\mathbf{k-q,}\uparrow }^{\dagger }a_{-\mathbf{k}%
\downarrow }\right\rangle $ of the dual Hamiltonian. The peak of the
correlation function in Eq. (6) at the relative momentum $\mathbf{-q}$ thus
confirms Bose condensation to a non-zero pair momentum for the original
Hamiltonian, and the distribution in $\mathbf{k}$ of the correlation $%
\left\langle a_{-\mathbf{k-q,}\uparrow }^{\dagger }a_{-\mathbf{k}\downarrow
}\right\rangle $ gives the original pair wavefunction. (iii) Similar to the
LOFF\ state, for a d-wave superfluid phase with the order parameter $%
\left\langle a_{\mathbf{k,}\uparrow }a_{-\mathbf{k}\downarrow }\right\rangle
\propto \cos k_{x}-\cos k_{y}$, , the pair wavefunction and its spatial
symmetry can be directly measured by detecting the correlation (6) for the
dual Hamiltonian.

\begin{figure}[tbph]
\centering
\includegraphics[height=3cm,width=8cm]{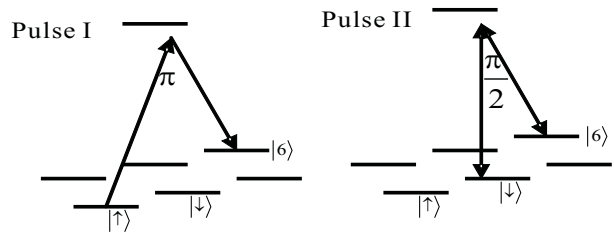}
\caption{ Illustration of the two Raman pulses (a $\protect\pi$ and a $%
\protect\pi/2$ pulses, respectively) before the time-of-flight for
measuring the correlation function in Eq. (6). We use the level
structure of $^6$Li atoms as an example (with the magnetic field
near the wide Feshbach resonance).} \label{Fig1}
\end{figure}

In summary, we have established the results as we outlined in the
introduction.

This work was supported by the MURI, the DARPA, the NSF award (0431476), the
DTO under ARO contracts, and the A. P. Sloan Fellowship.


\begin{thebibliography}{99}
\bibitem{note1}  For a plaquette configuration, the levels $\left| \uparrow
\right\rangle $ and $\left| \downarrow \right\rangle $ are actually doubly
degenerate due to the rotational symmetry (see \cite{11,12}), but the
physical picture is basically the same by adding a pseudospin index for this
degeneracy.
\end{thebibliography}
\end{document}